\RequirePackage{amsmath}
\documentclass[12pt]{iopart}
\usepackage{subfigure}
\usepackage{graphicx}
\usepackage{amsmath}
\usepackage{amsfonts}
\usepackage{mathtools}
\usepackage{amssymb}
\usepackage{xcolor}

\usepackage[utf8]{inputenc}

\usepackage{newtxtext}
\usepackage{newtxmath}

\usepackage{cite}
\DeclareMathAlphabet{\mathcal}{OMS}{cmsy}{m}{n}

\begin{document}

\title{Bounds on current fluctuations in periodically driven systems}
\author{Andre C Barato$^1$, Raphael Chetrite$^{2}$, Alessandra Faggionato$^{3}$, and Davide Gabrielli$^{4}$}
\address{
$^1$ Max Planck Institute for the Physics of Complex Systems,\\
N\"othnitzer Str. 38, 01187 Dresden, Germany\\
$^2$ Laboratoire J A Dieudonn\'e,\\
UMR CNRS 7351, Universit\'e de Nice Sophia Antipolis, Nice 06108, France\\
$^3$ Dipartimento di Matematica, Universit\`a di Roma ‘La Sapienza’,\\
P.le Aldo Moro 2, 00185 Roma, Italy\\
$^4$  DISIM, Universit\`a dell'Aquila\\
Via Vetoio,  Loc. Coppito, 67100 L'Aquila, Italy
}

\def\ex#1{\langle #1 \rangle}
\begin{abstract}
Small nonequilibrium systems in contact with a heat bath can be analyzed with the framework of stochastic thermodynamics.
In such systems, fluctuations, which are not negligible, follow universal relations such as the fluctuation theorem. More recently,
it has been found that, for nonequilibrium stationary states, the full spectrum of fluctuations of any thermodynamic current
is bounded by the average rate of entropy production and the average current. However, this bound does not apply to periodically driven systems, such as heat 
engines driven by periodic variation of the temperature and artificial molecular pumps driven by an external protocol. We obtain 
a universal bound on current fluctuations for periodically driven systems. This bound is a generalization of the known bound
for stationary states. In general, 
the average rate that bounds fluctuations in periodically driven systems is different from the rate of entropy production. We also 
obtain a local bound on fluctuations that leads to a trade-off relation between speed and precision in periodically driven systems, which 
constitutes a generalization to periodically driven systems of the so called thermodynamic uncertainty relation. From a technical 
perspective, our results are obtained with the use of a recently developed theory for 2.5 large deviations for  Markov jump processes 
with time-periodic transition rates.

\end{abstract}

\section{Introduction}

Thermodynamics \cite{callen} is a major branch of physics concerned with the limits of operation of machines that 
transform heat into other forms of energy. This theory is limited to macroscopic systems such as a 
steam engine. However, the way heat and temperature relate to other forms of energy is also   
important for small nonequilibrium systems, such as molecular motors and colloidal heat engines. 
For such systems, thermal fluctuations are relatively large and they cannot be ignored.

Stochastic thermodynamics \cite{seif12} generalizes thermodynamics to small nonequilibrium systems. A major question
that arises within this theoretical framework that takes fluctuations into account is: What are the universal relations that rule 
fluctuations in small nonequilibrium systems? The fluctuation theorem is one such relation \cite{evan93,gall95,jarz97,croo99,lebo99,seif05}, it is a constraint on the probability 
distribution of entropy that generalizes the second law of thermodynamics.  

A more recent universal relation associated with such fluctuations is the thermodynamic uncertainty relation from \cite{bara15a}. This relation establishes 
that precision of a thermodynamic current, such as the number of consumed ATP or the displacement of a molecular motor, has a minimal universal 
energetic cost. Possible applications of the thermodynamic uncertainty relation  include the inference of enzymatic schemes in single molecule 
experiments \cite{bara15}, a bound on the efficiency of molecular motors that depends only on fluctuations of the 
displacement of the motor \cite{piet16b}, a universal relation between power and efficiency for heat engines
in a stationary state \cite{piet18}, and design principles in nonequilibrium self-assembly \cite{nguy16}.

The thermodynamic uncertainty relation is a consequence of a more general bound on the full spectrum of current fluctuations \cite{piet16,ging16}. Using 
large deviation theory \cite{elli85,demb98,holl00,touc09}, this bound is expressed as a parabola that is above the so called rate function, which quantifies the
rate of exponentially rare  events. A key feature of this parabolic bound is that it depends solely on the average entropy production and the average current, i.e.,
knowledge of the average entropy production and the average current implies a bound on arbitrary fluctuations of any thermodynamic current. 
There has been much recent work related to this universal principle about current fluctuations 
\cite{piet16a,pole16,ray16,nyaw16,guio16,piet17b,horo17,pigo17,garr17,proe17,maes17,hyeo17,ging17a,bisk17,dech17,bran18,nard18,chiu18}.

The  parabolic bound applies to stationary states of Markov processes with time-independent transition rates.
Physically, this situation corresponds to systems that are driven by fixed thermodynamic forces, e.g., molecular 
motors driven by the free energy of ATP hydrolisis. Another  major class of thermodynamic systems
away from equilibrium is that of periodically driven systems, which can be described as  Markov processes with time-periodic 
transition rates. Two experimental realizations of periodically driven systems are Brownian heat engines \cite{mart17} and artificial molecular pumps \cite{erba15}. 

There is a fundamental difference with respect to fluctuations between systems driven by a fixed thermodynamic force and 
periodically driven systems. As shown in \cite{bara16}, for a periodically driven system, the energetic cost of precision 
of a thermodynamic current can be arbitrarily small, in stark contrast to systems driven by a fixed thermodynamic 
force, for which this precision has a minimal universal cost, as determined by the thermodynamic uncertainty relation. 
Hence, the parabolic bound from \cite{piet16,ging16} that depends on the average rate of entropy production does not apply to periodically driven systems. 
For the particular case of a time-symmetric protocol, a derivation of a thermodynamic uncertainty relation has been proposed in \cite{proe17}.
The relation between these two classes of nonequilibrium systems is also relevant for the mapping of artificial molecular machines, 
which are often driven by an external periodic protocol (see \cite{wils16} for a 
counter-example), onto biological molecular motors, which are autonomous machines driven by ATP, as discussed in \cite{raz16,rots17}. 

In this paper, we obtain a universal bound on current fluctuations in periodically driven systems that is also parabolic. For the particular 
case of a current with increments that do not depend on time, such as internal net motion in a molecular pump,  our bound depends on a single average rate. 
However, this average rate is different from the entropy production. For a constant protocol that leads to time-independent transition rates, our bound becomes an even more general bound than the 
known bound for stationary states from \cite{piet16,ging16}. A relevant technical aspect of our proof is as follows. The parabolic bound for stationary states has been proved in \cite{ging16}. 
This proof uses a remarkable result for large deviations in Markov processes, i.e., the exact form of the rate function for 2.5 large deviations for stationary states 
\cite{maes08,bara15c,bert15bis,bert15}. More recently,  the rate function of 2.5 large deviations for time-periodic transition rates has been obtained in \cite{bert18}. 
We use this result to prove our bounds. 
 
Similar to the parabolic bound for stationary states that implies the thermodynamic uncertainty relation, our global bound on large deviations leads to a 
trade-off relation between speed and precision in periodically driven systems. We obtain a tighter local bound on the rate function that leads 
to an improved trade-off relation between speed and precision. For the case of stationary 
states, this bound is also tighter then the bound determined by the thermodynamic uncertainty relation. 

We also prove our results for the case of a cyclic stochastic protocol \cite{bara16,ray17,bara18}. Such protocols are convenient to perform illustrative calculations 
with specific models. Furthermore, the proofs for stochastic protocols are a generalization of our results for deterministic protocols, since current 
fluctuations for a stochastic protocol with an infinitely large number of jumps are equivalent to current fluctuations for a deterministic protocol \cite{bara18}.    

The paper is organized in the following way. In Sec. \ref{sec2} we define the basic mathematical quantities and physical models. In Sec. \ref{sec3}, we introduce 
and illustrate our main results for the case of currents with time-independent increments. 
The bounds are derived in Sec. \ref{sec4}. We conclude in Sec. \ref{sec5}. \ref{appa} contains the proofs for the case of a stochastic protocol.

\section{Mathematical preliminaries and physical models}
\label{sec2}
\subsection{Markov processes with time-periodic transition rates and fluctuating observables}

We consider a Markov jump process with finite number of states  $\Omega$. The space of states is  written as $\{1,2,\dots, \Omega\}$.
The transition rate from state $i$ to state $j$ at time $t$ is denoted by $w_{ij}(t)$. Since we are interested in periodically driven systems, 
these transition rates have a period $\tau$, i.e., $w_{ij}(t)=w_{ij}(t+\tau)$. Furthermore, we assume that if $w_{ij}(t)\neq 0$ then $w_{ji}(t)\neq 0$. 

The master equation that governs the time-evolution of $P_i(t)$, the probability to be
in state $i$ at time $t$, reads 
\begin{equation}
\frac{d}{dt}P_i(t)=\sum_{j\neq i}\left[P_j(t)w_{ji}(t)-P_i(t)w_{ij}(t)\right].
\label{mastereq}
\end{equation}
In the long time limit, $P_i(t)$ tends to an invariant time-periodic distribution $\pi_i(t)=\pi_i(t+\tau)$. An important quantity in this paper is 
the average elementary current   
\begin{equation}
\mathcal{J}_{ij}(t)\equiv \pi_i(t)w_{ij}(t)- \pi_j(t)w_{ji}(t).
\label{avgelementary}
\end{equation}

Fluctuations can be analyzed if we consider stochastic variables that are defined as functionals of 
a stochastic trajectory $( a_t )_{0\leq t \leq m\tau}$, where $m\tau$ is the final time and $m$ is an integer number. This trajectory is a sequence of 
jumps and waiting times. If a jump takes place at time $t$, the state of the system before and after the jump 
is denoted by $a_t^-$ and $a_t^+$, respectively.
Two basic fluctuating  quantities
are    
\begin{equation}
\rho_{i}^{(m)}(t)\equiv \frac{1}{m}\sum_{k=0}^{m-1}\delta_{a_{\tau k+t},i}
\label{defden}
\end{equation}
and
\begin{equation}
C_{ij}^{(m)}(t)\equiv \frac{1}{m dt}\sum_{k=0}^{m-1}\left(\sum_{t'\in[t,t+dt]}\delta_{a_{\tau k+t'}^-,i}\delta_{a_{\tau k+t'}^+,j}\right),
\label{defflow}
\end{equation}
where $dt$ is an infinitesimal time-interval and $t\in [0,\tau]$. The empirical density $\rho_{i}^{(m)}(t)$ counts the fraction of periods with the system in state 
$i$ at time $t$. The empirical flow $C_{ij}^{(m)}(t)$ counts the number of jumps  per period  from 
$i$ to $j$ at  time  $t$. 
Even though both quantities are functionals of the 
stochastic trajectory, to simplify notation, we do not keep the explicit dependence on $( a_t )_{0\leq t \leq m\tau}$.  
The fluctuating empirical current from state $i$ to state $j$ is given by 
\begin{equation}
J^{(m)}_{ij}(t)= C^{(m)}_{ij}(t)-C^{(m)}_{ji}(t).
\label{defcur}
\end{equation} 
The average in Eq. \eqref{avgelementary} is   
\begin{equation}
\mathcal{J}_{ij}(t)=\lim_{m\to \infty} \langle J_{ij}^{(m)}(t)\rangle,
\end{equation} 
where the brackets denote an average over stochastic trajectories.

A generic current $J^{(m)}_\alpha$ is defined by its periodic increments $\alpha_{ij}(t)$, which are anti-symmetric, i.e.,   $\alpha_{ij}(t)=-\alpha_{ji}(t)$, as 
\begin{equation}
J^{(m)}_\alpha \equiv \frac{1}{\tau}\int_{0}^{\tau}dt\sum_{i<j}\alpha_{ij}(t)J_{ij}^{(m)}(t),
\label{defjalpha}
\end{equation}
where $\sum_{i<j}$ represents a sum over all pairs of states $(i,j)$ with $i<j$ and with  non-zero transition rates. 
The current in Eq. \eqref{defjalpha} can also be written in the form
\begin{equation}
J_{\alpha}^{(m)}=\frac{1}{m\tau}\sum_{\substack{s\in\left[0,m\tau\right]:\\ a_{s}^{-}\neq a_{s}^{+}}}\alpha_{a_{s}^{-} a_{s}^{+}}(s).
\label{defjalpha2}
\end{equation}
In stochastic thermodynamics, physical observables such as heat fluxes and particle fluxes are expressed as currents $J_\alpha^{(m)}$.
The average rate associated with $J^{(m)}_\alpha$ in the limit $m\to \infty$ reads  
\begin{equation}\label{pranzo}
\mathcal{J}_\alpha\equiv \lim_{m\to \infty}\langle J^{(m)}_\alpha\rangle=\frac{1}{\tau}\int_{0}^\tau dt\sum_{i<j}\alpha_{ij}(t)\mathcal{J}_{ij}(t).
\end{equation}
Furthermore, the diffusion coefficient associated with $J^{(m)}_\alpha$  is defined as 
\begin{equation}
D_\alpha\equiv \lim_{m\to \infty}m\tau \frac{\langle (J^{(m)}_\alpha-\mathcal{J}_\alpha)^2\rangle}{2}.
\label{diffdef}
\end{equation}

An important current in stochastic thermodynamics is the entropy increase of the environment  \cite{seif12}, which corresponds to the
increments  $\alpha_{ij}(t)=\ln\frac{w_{ij}(t)}{w_{ji}(t)}$. The average rate of entropy production is then given by 
\begin{equation}
\sigma\equiv\frac{1}{\tau}\int_{0}^\tau dt\sum_{i<j}\ln\frac{w_{ij}(t)}{w_{ji}(t)}\mathcal{J}_{ij}(t)=\frac{1}{\tau}\int_{0}^\tau dt\sum_{i<j}\ln\frac{\pi_i(t)w_{ij}(t)}{\pi_j(t)w_{ji}(t)}\mathcal{J}_{ij}(t).
\label{entrate} 
\end{equation}
The second equality follows from $\pi_i(t)=\pi_i(t+\tau)$ and from Eq. \eqref{mastereq}, which leads to   $\partial _t \pi_i + \sum_{j\neq i} \mathcal{J}_{ij} =0$.

\subsection{Large deviations}

The rate function from large deviation theory quantifies exponentially rare events in 
the long time limit \cite{elli85,demb98,holl00,touc09}. It is defined through  the relation
\begin{equation}
	\textrm{Prob}(J^{(m)}_\alpha\approx x)\sim \exp[-m\tau I_\alpha(x)],
\end{equation}
where the symbol $\sim$ means asymptotic equality in the limit $m\to\infty$ and $J^{(m)}_\alpha\approx x$ means 
that  $J^{(m)}_\alpha$ lies in an infinitesimal interval around $x$. Our main result  is a parabola that bounds $I_\alpha(x)$, which is a convex function, from above. This parabola depends on an  
average rate. For the known parabolic bound for stationary states from \cite{piet16,ging16}, this rate is the average rate of entropy production $\sigma$ in Eq. \eqref{entrate}. In our bound
for periodically driven systems, this rate is, in general, different from $\sigma$.

Current fluctuations can also be characterized by the scaled cumulant generating function 
\begin{equation}
	\lambda_\alpha(z)\equiv \lim_{m\to \infty}\frac{1}{m\tau}\ln\langle\exp(m\tau J_\alpha^{(m)}z)\rangle,
	\label{defscaled}
\end{equation}
where $z$ is a real number.
The cumulants associated with $J_\alpha^{(m)}$ can be obtained as derivatives of $\lambda_\alpha(z)$ at $z=0$. 
The scaled current generating function $\lambda_\alpha(z)$ is a Legendre-Fenchel transform of the rate function $I_\alpha(x)$, i.e.
\begin{equation}
	\lambda_\alpha(z)\equiv \sup_{x}\{xz-I_\alpha(x)\}.
	\label{letransf}
\end{equation}
If a parabola bounds $I_\alpha(x)$ from above then a corresponding parabola, which can be determined from Eq. \eqref{letransf}, bounds $\lambda_\alpha(z)$ from below.
For illustrations of our results we perform calculations of $\lambda_\alpha(z)$ using  known methods \cite{bara16,bara18}.   

\subsection{Stochastic protocol}\label{lego}

We also consider the case of an external protocol that is stochastic \cite{bara16,ray17,bara18}. In order to mimick a deterministic periodic protocol, this stochastic protocol is cyclic and has $N$ states. The transition rate from 
state $i$ to state $j$ with the external protocol in state $n=0,1,\ldots,N-1$ is denoted by $w_{ij}^n$. The transition rate for a change in the external protocol from state $n$ to state $n+1\mod{N}$
is $\gamma$, whereas the transition rate for the reversed transition is $0$. Consider 
a deterministic periodic protocol characterized by the rates   $w_{ij}(t)$ and 
the period $\tau$. If the rates of the model with a stochastic protocol are $w_{ij}^n= w_{ij}(t=n\tau/N)$ and $\gamma= N/\tau$, then in the limit of $N\to\infty$, current fluctuations for the stochastic 
protocol become equal to current fluctuations for the deterministic protocol \cite{bara18}. Hence, the deterministic protocol corresponds to an  asymptotic limit of  a stochastic protocol. We point 
out that we do not consider the cost of the external protocol \cite{bara17aa}.

In \ref{appa}, we derive bounds on current fluctuations  for the case of a stochastic protocol. These derivations are similar to the derivation in Sec. \ref{sec4} for a deterministic periodic protocol.
An advantage of models with a stochastic protocol is that they are Markov processes with time-independent transition rates, which can simplify the exact evaluation 
of the scaled cumulant generating function in Eq. \eqref{defscaled}, as explained in \cite{bara16}. Whereas the expressions in the main text are for the case of 
a deterministic protocol, the expressions for a stochastic protocol can be obtained 
from these expressions for a deterministic protocol by making the substitution $\tau^{-1}\int_0^\tau dt\to N^{-1}\sum_n$, as explained in  \ref{appa}.

\subsection{Case studies}

\subsubsection{Colloidal particle driven by a time-periodic field}\label{sec241}
\begin{figure}
\subfigure[]{\includegraphics[width=50mm]{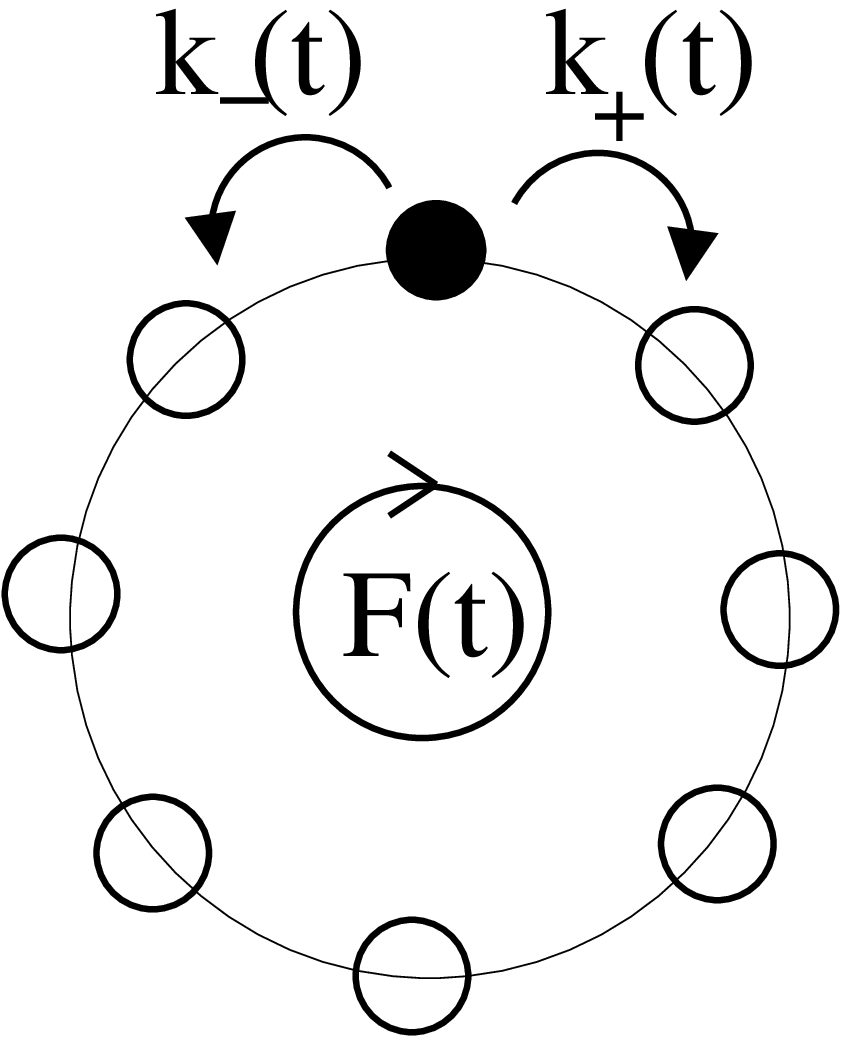}\label{fig1a}}
\subfigure[]{\includegraphics[width=50mm]{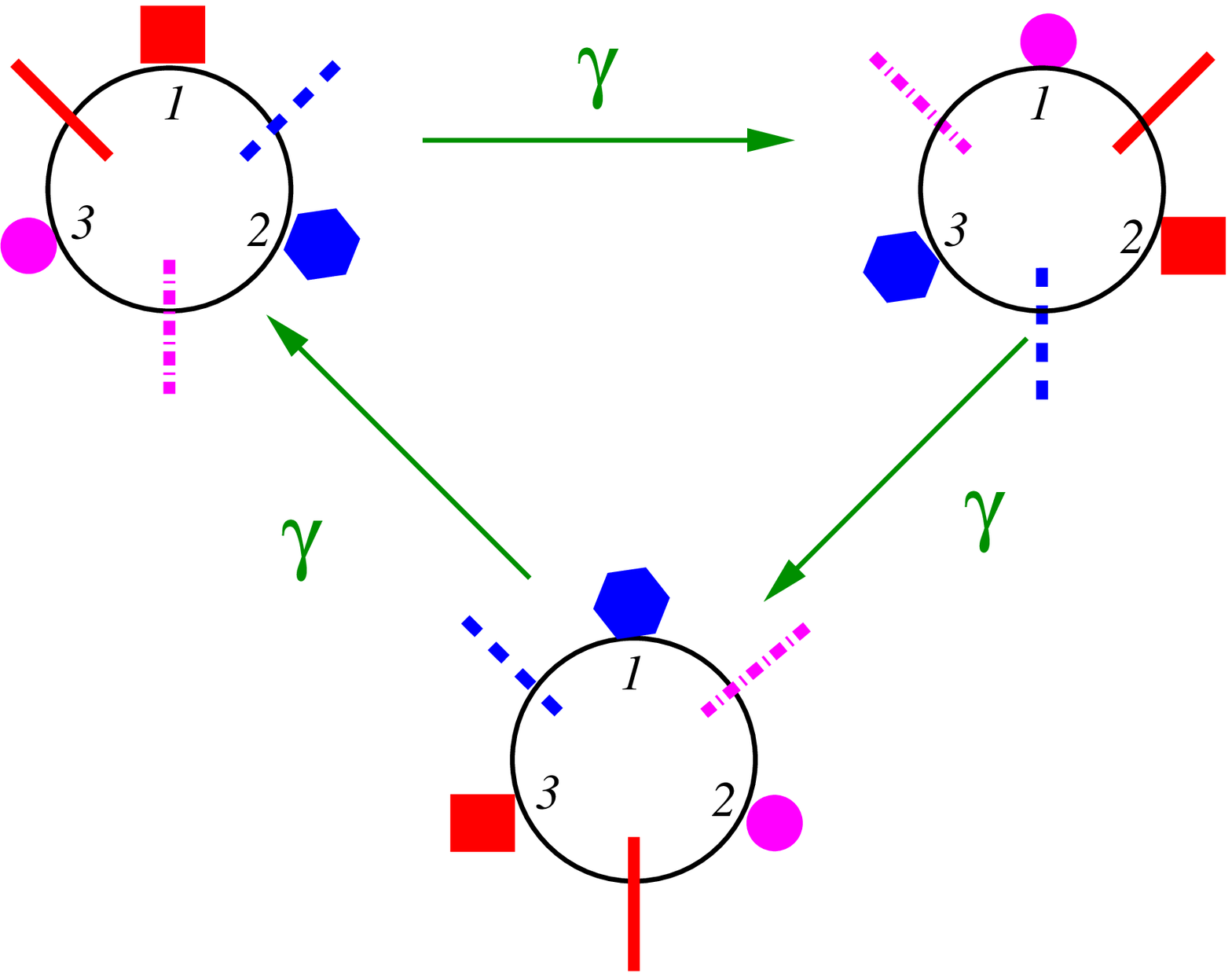}\label{fig1b}}
\subfigure[]{\includegraphics[width=50mm]{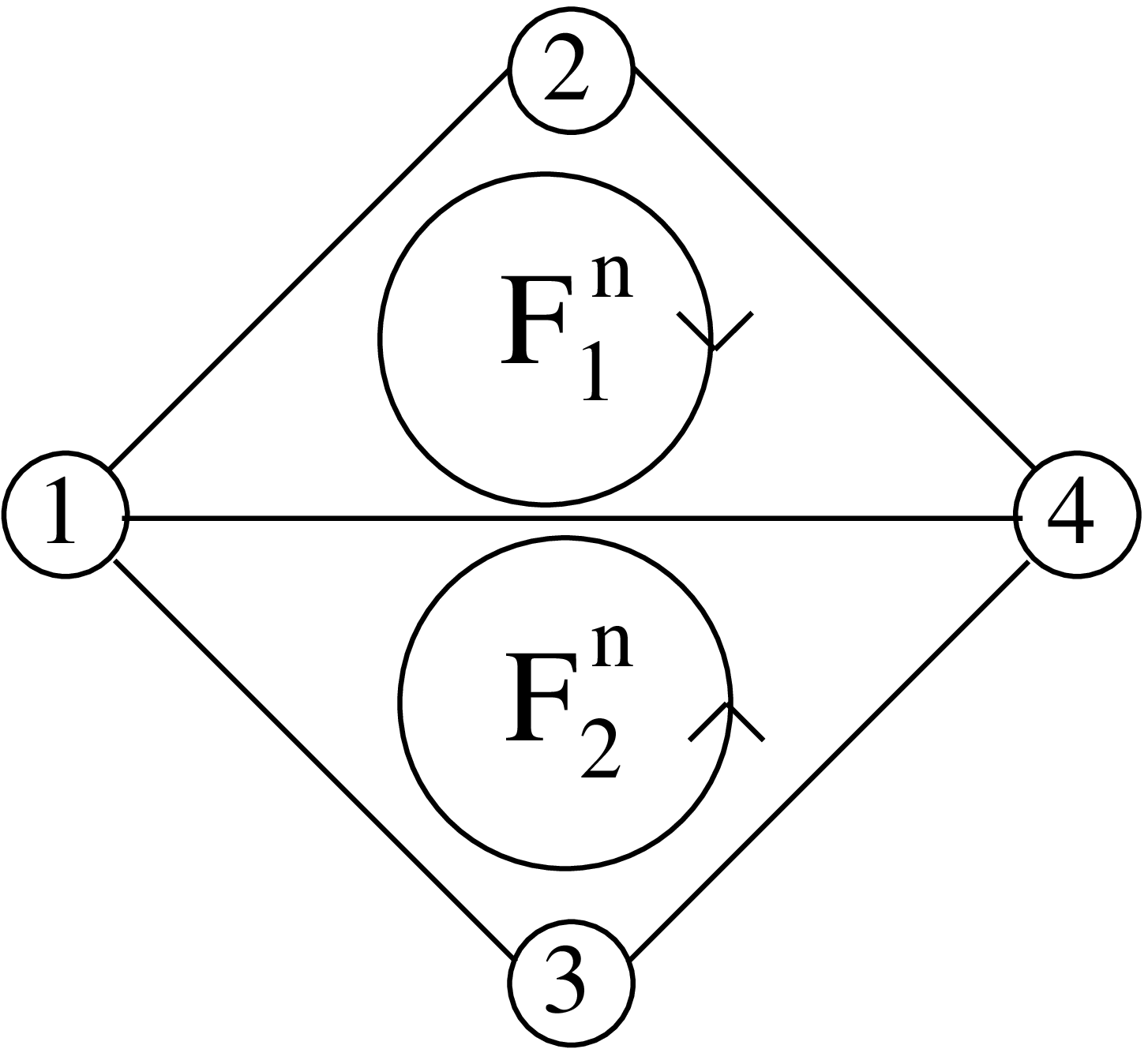}\label{fig1c}}
\vspace{-3mm}
\caption{Case studies. (a) Biased random walk with time-periodic force $F(t)$. (b) Model for a molecular pump. The red square represents the energy $E_1$, the blue hexagon represents 
the energy $E_2$, and the magenta circle represents the energy $E_3$. The red solid bar represents the energy barrier $B_1$, the blue dashed bar represents the energy barrier $B_2$, and 
the dotted magenta bar represents the energy barrier $B_3$. The green arrows represent transitions that change the state of the protocol.  (c) Representation of the network of states of the 
model with 4 states and two independent thermodynamic forces that depend on the state of the external protocol $n$.
}
\label{fig1}
\end{figure}

The first model in Fig. \ref{fig1a} is a biased random walk on a ring with $\Omega$ states driven by a time-periodic force 
$F(t)\equiv F_0 \cos(2\pi t/\tau)$.
A physical realization of this model is a charged colloid on a ring subjected to a time-periodic electrical field. We set Boltzmann constant $k_B$ and the temperature $T$ to $k_BT=1$ throughout.
The transition rate for a jump in the clockwise direction is $k_+(t)\equiv k\textrm{e}^{F(t)/\Omega}$ and the reversed transition rate is   $k_-(t)\equiv k$. These transition rates satisfy the generalized detailed
balance relation  \cite{seif12}. The current we consider is the net number of jumps in the clockwise direction per unit time. For this model, the scaled cumulant generating function in Eq. \eqref{defscaled} can 
be calculated exactly \cite{bara18}.

\subsubsection{Molecular pump}\label{sec242}
The other two models are driven by a stochastic protocol.  The model illustrated in Fig. \ref{fig1b} is a molecular pump with $\Omega=3$. This 
model has been introduced in \cite{bara16}. The external protocol changes energies and energy barriers between states, which can 
lead to net rotation in the ring with three states. The number of states of the external protocol is $N=3$. The states of the external protocol  are denoted by $0,1,2$, which correspond respectively  to the top left circle, the 
top right circle and the bottom circle  in  Fig. \ref{fig1b}. In this model, the energies and energy barriers are rotated in the clockwise direction by one step if 
a jump (with rate $\gamma$ ) that changes the state of the protocol takes place. The energies are denoted by $E_1$, $E_2$, and $E_3$, whereas the energy barriers are denoted by $B_1$, $B_2$, and $B_3$. 
The internal transition rates  are given by
\begin{equation}
w_{ij}^n= \textrm{e}^{E_{i-n}-B_{j-n}},
\end{equation}
for $j=i+1$, and 
\begin{equation}
w_{ij}^n= \textrm{e}^{E_{i-n}-B_{i-n}},
\end{equation}
for $j=i-1$, where we assume periodic boundary conditions. 
An important property of molecular pumps is that the thermodynamic force is zero for any state $n$ of the external protocol. This physical condition is manifested in the following restriction on the transition rates 
\begin{equation}
\frac{w_{12}^nw_{23}^nw_{31}^n}{w_{21}^nw_{32}^nw_{13}^n}=1.
\end{equation}
The current we consider is the net number of jumps in the clockwise direction per unit time.  
The scaled cumulant generating function in Eq. \eqref{defscaled} associated with this current can be calculated from the eigenvalue  
of a  modified generator, as shown in \cite{bara16}. 

\subsubsection{Enzymatic reaction with stochastic substrate concentrations}\label{sec243}
The model illustrated in Fig. \ref{fig1c} is a model with $\Omega=4$ and two independent  thermodynamic forces $F_1^n$ and $F_2^n$, which depend on the state of the external protocol $n$. This model 
can be interpreted as a enzyme that can consume two different substrates and produces one product \cite{bara15a}. The two enzymatic cycles are $E+S_1\to ES_1\to EP\to E+P$ and  $E+S_2\to ES_2\to EP\to E+P$,
where $E$ is the enzyme, $P$ is the product, $S_1$ is one substrate, and $S_2$ is another substrate. State $1$ corresponds to the free enzyme $E$, state $2$ corresponds to $ES_1$, 
state $3$ corresponds to $ES_2$, and state $4$ corresponds to $EP$. The external control of the concentrations of the substrates $S_1$ and $S_2$ generate 
thermodynamic forces that depend on $n$. The number of states of the external protocol is $N=2$. The generalized detailed balance relation for this model reads 
\begin{equation}
F_1^n= \ln\frac{w^n_{12}w^n_{24}w^n_{41}}{w^n_{21}w^n_{42}w^n_{14}}\qquad F_2^n= \ln\frac{w^n_{13}w^n_{34}w^n_{41}}{w^n_{31}w^n_{43}w^n_{14}}.
\end{equation}
The thermodynamic forces change between two values of the same modulus and different sign stochastically, i.e., $F_1^n$ is given by $F_1^0=F_1$ and  $F_1^1=-F_1$,  whereas  $F_2^n$ is given by $F_2^0=F_2$ and  $F_2^1=-F_2$. 

The transition rate for a change of 
the external protocol is $\gamma$. The transitions rates are set to 
$w_{12}^n= k \textrm{e}^{F_1^n/2}$, $w_{13}^n= k \textrm{e}^{F_2^n/2}$, $w_{14}^n= k$,  $w_{21}^n= k$, $w_{24}^n= k \textrm{e}^{F_1^n/2}$, $w_{31}^n= k$, $w_{34}^n= k \textrm{e}^{F_2^n/2}$,
$w_{41}^n= k$, $w_{42}^n= k$, and $w_{43}^n= k$. The current we consider is the elementary current from state $1$ to state $2$, which corresponds to the net number of $S_1$ molecules that have been consumed per 
unit time. As is the case of the previous model, the scaled cumulant generating function in Eq. \eqref{defscaled} can be calculated with the method explained in \cite{bara16}.

\section{Main results}
\label{sec3}

In this Section we discuss our main results for currents with time-independent increments $\alpha_{ij}(t)=\alpha_{ij}$, 
which include the case of currents generated in a molecular pump. For time-independent increments, the results acquire a simpler 
form with a more direct physical interpretation. In Sec. \ref{sec4}, we present proofs of more general results, which, {\sl inter alia}, also hold for  
currents with time-dependent increments. Physical examples of currents with time-dependent increments include the heat and work 
currents in heat engines (see \cite{ray17} for general definitions of these currents). The general features of our main results 
presented in this Section are the same irrespective of whether 
the protocol is deterministic or stochastic, which is discussed in \ref{appa}.

\subsection{Global bound}

\begin{figure}
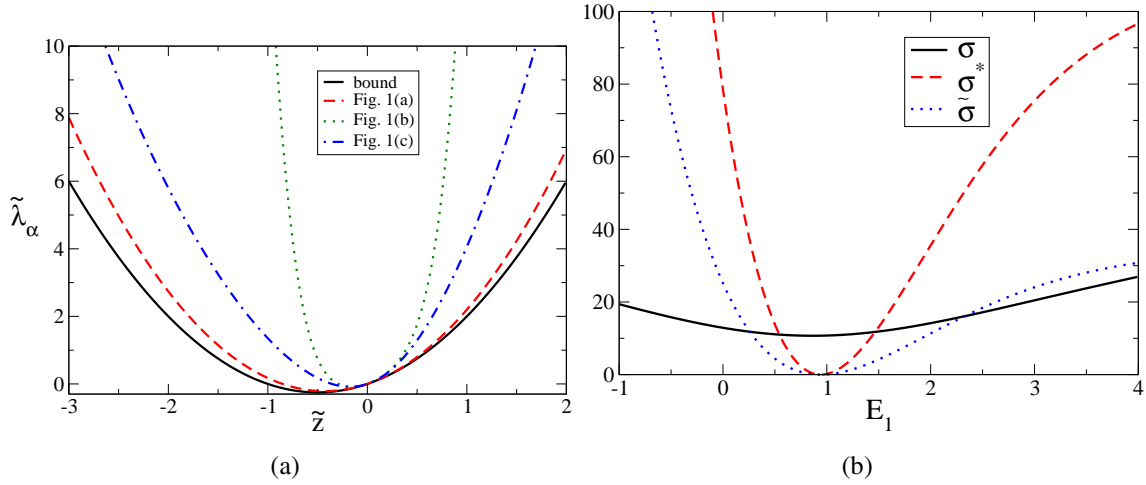

\subfigure[]{\includegraphics[width=75mm]{fig2a.eps}\label{fig2a}}
\subfigure[]{\includegraphics[width=75mm]{fig2b.eps}\label{fig2b}}
\vspace{-3mm}
\caption{Illustration of the bound. (a) The function $\tilde{\lambda}_\alpha(\tilde{z})$ in Eq. \eqref{scalingfunction} for the models 
from Fig. \ref{fig1}, as indicated in the legends, compared to the lower bound $\tilde{z}(1+\tilde{z})$. The parameters for the 
model represented in Fig \ref{fig1a} are set to $F_0/\Omega=2$ and $k=\tau=1$. The parameters for the 
model represented in Fig \ref{fig1b} are set to $E_1= E_3= B_1=B_2=0$, $E_2=2$, $B_3=5$,  and $\gamma=1/10$. The parameters for the 
model represented in Fig. \ref{fig1c} are set to $F_1=2$, $F_2=1/2$, $k=1$, and $\gamma=1/10$. (b) Comparison between the rate of entropy 
production $\sigma$, the rate $\sigma^*$ and the rate $\tilde{\sigma}$,  for the model in Fig. \ref{fig1b} with parameters $E_2=2$, $E_3=-5$, $B_1=-5$, $B_2=2$, $B_3=0$, and $\gamma=\textrm{e}^{2}$.
The parameter $E_1$ is the variable in the horizontal axis.
}
\label{fig2}
\end{figure}

The parabolic bound on the rate function is
\begin{equation}
I_\alpha(x)\le \frac{\sigma^*}{4\mathcal{J}_\alpha^2}(x-\mathcal{J}_\alpha)^2, 
\label{mainresultspe1}
\end{equation}
where 
\begin{equation}
\sigma^*\equiv \frac{1}{\tau}\int_0^{\tau}\sum_{i<j}\frac{(\bar{\mathcal{J}}_{ij})^2}{\mathcal{J}_{ij}(t)}\ln\frac{\pi_i(t)w_{ij}(t)}{\pi_j(t)w_{ji}(t)}dt,
\label{defsigmastar} 
\end{equation}
and
\begin{equation}
\bar{\mathcal{J}}_{ij}\equiv \frac{1}{\tau}\int_0^{\tau}\mathcal{J}_{ij}(t)dt.
\label{defcurrtime}
\end{equation}
The inequality $\sigma^*\ge 0$ comes from the fact that for fixed $t$ every term in the sum $\sum_{i<j}$ in Eq. \eqref{defsigmastar} is not negative. 
In general, the average rate $\sigma^*$ is different from the thermodynamic rate of entropy production $\sigma$ in Eq. \eqref{entrate}. Furthermore, 
there is no simple inequality relating both quantities, as illustrated in Fig. \ref{fig2b}.

For the case of time-independent transition rates $w_{ij}(t)= w_{ij}$, $\sigma^*= \sigma$  and the bound \eqref{mainresultspe1} becomes 
\begin{equation}
I_\alpha(x)\le \frac{\sigma}{4\mathcal{J}_\alpha^2}(x-\mathcal{J}_\alpha)^2. 
\end{equation}
This bound is the known parabolic bound for time-independent transition rates proved in \cite{ging16}. Hence, Eq. \eqref{mainresultspe1}  constitutes 
a generalization of this parabolic bound to periodically driven systems.

In terms of the scaled cumulant generating function, the bound in Eq. \eqref{mainresultspe1} is written as 
\begin{equation}
\lambda_\alpha(z) \ge z \mathcal{J}_\alpha(1+ \mathcal{J}_\alpha z/\sigma^*),
\label{mainresultspe2}
\end{equation}
where we used Eq. \eqref{letransf}. The universality of our result is illustrated in Fig. \ref{fig2a}. There we compare the function
\begin{equation}
\tilde{\lambda}_\alpha(\tilde{z})\equiv \lambda_\alpha(z)/ \sigma^*=\lambda_\alpha(\tilde{z} \sigma^*/\mathcal{J}_\alpha)/ \sigma^*\ge \tilde{z}(1+\tilde{z}),
\label{scalingfunction}
\end{equation}
where $\tilde{z}\equiv z\mathcal{J}_\alpha/\sigma^{*}$, for the models in Fig. \ref{fig1}, with the lower bound $\tilde{z}(1+\tilde{z})$. 
This bound, or the bound in Eq. \eqref{mainresultspe1}, is a particular case of two bounds, one derived in Sec. \ref{roma} and the other derived in Sec. \ref{sec45}.

\subsection{Trade-off between speed and precision}

\begin{figure}
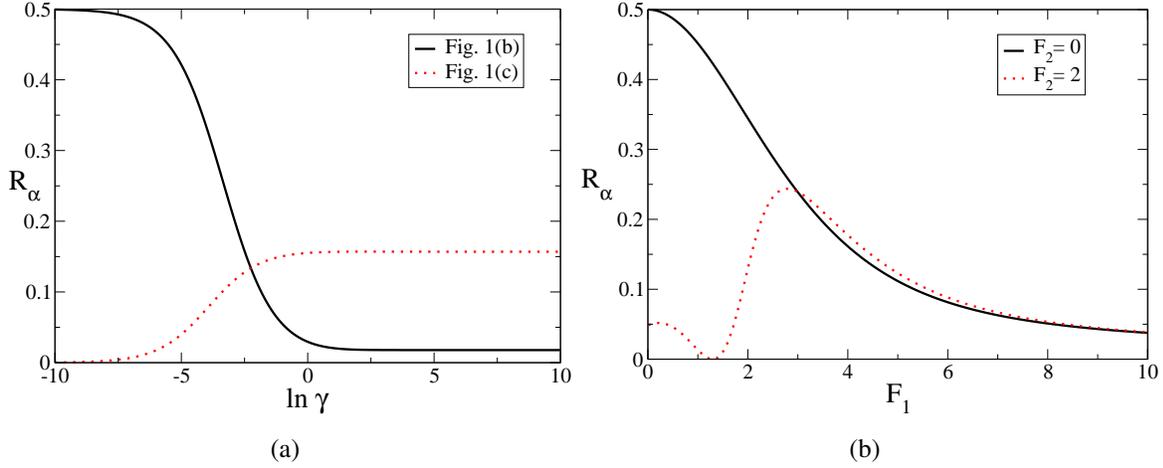

\subfigure[]{\includegraphics[width=75mm]{fig3a.eps}\label{fig3a}}
\subfigure[]{\includegraphics[width=77mm]{fig3b.eps}\label{fig3b}}
\vspace{-3mm}
\caption{Illustration of the trade-off relation. (a) The ratio $R_\alpha\equiv\mathcal{J}_\alpha^2/(2D_\alpha\sigma^*)\le 1/2$ 
as a function of the rate $\gamma$ for jump of the protocol. We have analyzed the model illustrated in Fig. \ref{fig1b} with parameters $E_1=1$, $B_1=5$, and $E_2=E_3=B_2=B_3=0$, 
and the model illustrated in Fig. \ref{fig1c} with parameters $F_1=F_2=k=1$. (b) The ratio $R_\alpha\equiv\mathcal{J}_\alpha^2/(2D_\alpha \sigma^*)\le 1/2$ 
as a function of $F_1$ Fig. \ref{fig1c} with parameters $k=\gamma=1$ and two values of $F_2$.
}
\label{fig3}
\end{figure}

Taking the second derivative of $I_\alpha(x)$ at $x=\mathcal{J}_\alpha$, we obtain the diffusion coefficient $D_\alpha$ 
defined in Eq. \eqref{diffdef} as
\begin{equation}
I''(\mathcal{J}_\alpha)= \frac{1}{2 D_\alpha}.
\label{DIsecond}
\end{equation}
The inequality in Eq. \eqref{mainresultspe1} and the fact that this inequality is saturated at $x=\mathcal{J}_\alpha$, leads to the following bound on $D_\alpha$,
\begin{equation}
D_\alpha\ge \frac{\mathcal{J}_\alpha^2}{\sigma^*}.
\end{equation}
In Sec. \ref{sec43}, we derive a local quadratic bound on $I_\alpha(x)$, which is valid for $x$ close to the average $\mathcal{J}_\alpha$. This local bound together with 
Eq. \eqref{DIsecond}, gives a tighter bound on $D_\alpha$ that reads   
\begin{equation}
D_\alpha\ge \frac{\mathcal{J}_\alpha^2}{\tilde{\sigma}}\ge \frac{\mathcal{J}_\alpha^2}{\sigma^*},
\label{boundD}
\end{equation}
where 
\begin{equation}
\tilde{\sigma}\equiv \frac{1}{\tau}\int_{0}^{\tau}dt\sum_{i<j}\frac{2(\bar{\mathcal{J}}_{ij})^2}{\pi_i(t) w_{ij}(t)+\pi_j(t) w_{ji}(t)}.
\label{sigmalocal}
\end{equation}
The second inequality in Eq. \eqref{boundD} is a consequence of $\sigma^*\ge \tilde{\sigma}$, which follows from the inequality 
\begin{equation}\label{basicine}
(a-b)\ln \frac{a}{b}\ge \frac{2(a-b)^2}{a+b},
\end{equation}
where $a$ and $b$ are positive. An inequality similar to $\sigma^*\ge \tilde{\sigma}$ has been considered in \cite{shira16}.
We point out that there is no general inequality between the entropy production $\sigma$ and 
the rate $\tilde{\sigma}$, as illustrated in Fig. \ref{fig2b}.

Rearranging the terms in Eq. \eqref{boundD}, we write the following universal trade-off 
relation between speed and precision for periodically driven systems,  
\begin{equation}
\mathcal{F}_\alpha^{-1}\mathcal{J}_\alpha\le \frac{\tilde{\sigma}}{2}\le  \frac{\sigma^*}{2}
\label{Fanotrade}
\end{equation}
where $\mathcal{F}_\alpha\equiv 2 D_\alpha/\mathcal{J}_\alpha$ is the Fano factor. The Fano factor 
characterizes the precision associated with $J_\alpha^{(m)}$, whereas $\mathcal{J}_\alpha$ quantifies the 
speed. In periodically driven systems, a current with small fluctuations, as characterized by a small Fano factor $\mathcal{F}_\alpha$, 
can only be as fast as $\tilde{\sigma}\mathcal{F}_\alpha/2$. 

This trade-off relation is a generalization of the thermodynamic uncertainty relation to periodically driven systems. In particular, for 
the case of time-independent transition rates $w_{ij}(t)= w_{ij}$, inequality \eqref{Fanotrade} implies the thermodynamic uncertainty relation 
$\mathcal{F}_\alpha^{-1}\mathcal{J}_\alpha\le \sigma/2$, since $\sigma^*=\sigma$ for this case. Furthermore, the inequality  
$\mathcal{F}_\alpha^{-1}\mathcal{J}_\alpha\le \tilde{\sigma}/2$, for time-independent transition rates, provides an even tighter bound than the 
thermodynamic uncertainty relation.

This result is relevant for the mapping between an artificial molecular 
pump and a system driven by a fixed thermodynamic force such as a biological molecular motor, which is modelled with time-independent transition rates that lead 
to a nonequilibrium stationary state, proposed in \cite{raz16}. 
With this mapping, one can construct a molecular pump that mimicks a stationary state and vice-versa, in the sense that both the average rate of entropy production and 
the average elementary currents between a pair of states are conserved. However, a mapping of a molecular pump onto a stationary state that also preserves fluctuations 
is not always possible, since a molecular pump may not fulfill   the relation $\mathcal{J}_\alpha^2/(2D_\alpha\sigma)\le 1/2$, as shown in \cite{bara16},
whereas a system that reaches a nonequilibrium stationary state must fulfill this relation. 

Our trade-off relations do not imply the generalization of the thermodynamic uncertainty relation from \cite{proe17} for the case of periodic protocols that are symmetric, i.e., 
$w(\tau/2+\Delta t)= w(\tau/2-\Delta t)$, where $0\le\Delta t\le \tau/2$. The trade-off relation from this reference involves the thermodynamic entropy production $\sigma$ and for
symmetric protocols the rate $\sigma$ is, in general, different from the rates  $\sigma^*$ and $\tilde{\sigma}$.

\subsection{Discussion of the bounds}

In Fig. \ref{fig3a}, we show plots of $R_\alpha\equiv \mathcal{J}_\alpha^2/(2D_\alpha\sigma^*)\le 1/2$ as a function of the rate $\gamma$, which quantifies 
the speed of the protocol, for the 
models illustrated in Fig. \ref{fig1b} and in Fig. \ref{fig1c}. For the first model, which is a molecular pump, we find that this bound is saturated 
if the transitions of the protocol are much slower than the internal transition rates associated with changes of the state of the system. For this model, 
in this limit the bound is saturated independent of the values of the energies and energy barriers. However, for the second model the bound is not saturated  in this limit.

In Fig. \ref{fig3b}, we show plots of $R_\alpha\equiv \mathcal{J}_\alpha^2/(2D_\alpha\sigma^*)\le 1/2$ for the model  illustrated in Fig. \ref{fig1c}. The quantity   
in the horizontal axis is the thermodynamic force $F_1$. For this model, the bound is saturated for $F_1$ small and the other thermodynamic force $F_2=0$. This 
saturation of the bound is similar to the saturation of the bound for stationary  states known as thermodynamic uncertainty relation, which happens in the linear response regime \cite{bara15a}.  

Let us comment on the rate $\sigma^*$ that we have introduced here. Its physical interpretation is that 
$\sigma^*$, and not the rate of entropy production $\sigma$, provides a bound on the whole spectrum of fluctuations for any current (with time-independent increments) 
in a generic periodically driven system arbitrarily far from equilibrium. In terms of the trade-off relation from Eq. \eqref{Fanotrade}, $\sigma^*$ (and also $\tilde{\sigma}$) 
provides a limit on how precise and fast a thermodynamic current can be. The rate of entropy production $\sigma$ quantifies the energetic 
cost of sustaining the operation of the nonequilibrium system. Interestingly, for time-independent transition rates corresponding to a system driven by a fixed thermodynamic force, 
$\sigma^*=\sigma$ is a rate that has both physical properties, i.e., it bounds current fluctuations and quantifies energetic cost.  

\subsection{$\sigma^*$ as the entropy production of a nonequilibrium stationary state}

The rate  $\sigma^*$ of the original periodically driven system can be interpreted as the rate of entropy production associated with the stationary state of an auxiliary Markov process with time-independent transition rates
that are determined by time-averaged quantities associated with the original system. These time-averaged quantities are $\bar{\mathcal{J}}_{ij}$, defined in Eq. \eqref{defcurrtime}, and 
\begin{equation}
\theta_{ij}\equiv  \frac{1}{\tau}\int_0^{\tau}\frac{\bar{\mathcal{J}}_{ij}}{\mathcal{J}_{ij}(t)}\ln\frac{\pi_i(t)w_{ij}(t)}{\pi_j(t)w_{ji}(t)}dt.
\label{defteta} 
\end{equation}
Both quantities are anti-symmetric, i.e., $\bar{\mathcal{J}}_{ij}=-\bar{\mathcal{J}}_{ji}$ and $\theta_{ij}=-\theta_{ji}$. Moreover, from the definition in Eq. \eqref{defteta}, 
$\bar{\mathcal{J}}_{ij}$ and $\theta_{ij}$ have the same sign. We assume without loss of generality that $\bar{\mathcal{J}}_{ij}$ and $\theta_{ij}$ are non-negative .

From Eq. \eqref{defsigmastar}, $\sigma^*$  can be written as $\sigma^*=\sum_{i<j}\bar{\mathcal{J}}_{ij}\theta_{ij}$. The transition rates associated with this 
auxiliary process are denoted by $r_{ij}$ and the stationary distribution associated with this 
process is denoted by $p_i$. The stationary probability currents of this auxiliary process are the time-averaged currents $\bar{\mathcal{J}}_{ij}$, hence, we have the constraint 
\begin{equation}
p_ir_{ij}-p_jr_{ji}= \bar{\mathcal{J}}_{ij}. 
\label{rconst1}
\end{equation}
Furthermore, if we impose  
\begin{equation}
\frac{r_{ij}}{r_{ji}}= \textrm{e}^{\theta_{ij}}, 
\label{rconst2}
\end{equation}
then the rate of entropy production of the auxiliary process is $\sigma^*$, i.e., $\sigma^*=\sum_{i<j}\bar{\mathcal{J}}_{ij}\ln(r_{ij}/r_{ji})$. From the conditions 
in Eq. \eqref{rconst1} and Eq. \eqref{rconst2}, we obtain
\begin{equation}
r_{ij}=\frac{ \bar{\mathcal{J}}_{ij}\textrm{e}^{\theta_{ij}}}{\textrm{e}^{\theta_{ij}}p_i-p_j}. 
\label{eqauxpro}
\end{equation}
The reversed rate $r_{ji}$ is then given by    
\begin{equation}
r_{ji}=\frac{ \bar{\mathcal{J}}_{ij}}{\textrm{e}^{\theta_{ij}}p_i-p_j}. 
\label{eqauxpro2}
\end{equation}
Equation \eqref{eqauxpro} defines a class of stationary states that have entropy production $\sigma^*$. Since transition rates are non-negative, the stationary probability 
must satisfy the constraint $\textrm{e}^{\theta_{ij}}p_i-p_j\ge 0$. One possible stationary probability that fulfills this constraint for any model is the uniform distribution 
$p_i=1/\Omega$ for $i=1,2,\ldots,\Omega$, since $\theta_{ij}\ge 0$.

We can now provide the following physical interpretation for $\sigma^*$. This rate quantifies the thermodynamic cost to maintain a non-equilibrium stationary state that is 
determined by the transition rates in Eq. \eqref{eqauxpro}. There are different stationary probabilities that fulfill Eq. \eqref{eqauxpro}, hence, this non-equilibrium 
stationary state is not unique but rather a class of nonequilibrium stationary states. The network topology of this class of nonequilibrium stationary states is the same as the 
network topology of of the periodically driven system, furthermore,  the stationary currents are the same as the time-averaged currents of the periodically driven system.
As an example, consider a colloidal particle  driven by an external periodic protocol, such as 
the model represented in Fig. \ref{fig1b}. For such molecular pump we can think of a colloidal particle driven by a fixed force that reaches a nonequilibrium 
stationary state. The force that drives this particle and the specific transition rates that determine its dynamics are obtained from time-averaged quantitative associated 
with the original molecular pump. The rate $\sigma^*$ quantifies the energetic cost of driving the colloidal particle with such fixed force.

\section{General bounds }
\label{sec4}

In this Section we derive the bounds that imply the results discussed in Sec. \ref{sec3}. 
We obtain two global bounds that imply the global bound in Eq. \eqref{mainresultspe1}, the first one is
given in Eq. \eqref{mainresult2} and the second one is given in Eq. \eqref{unbound}. 
We also  derive a local  bound that leads to the inequality in Eq. \eqref{boundD2}, which generalizes the 
trade-off relation in Eq. \eqref{Fanotrade}.

\subsection{First global bound}
\label{roma}

In our proof we use the theory for 2.5 large deviations for periodically driven systems  developed in \cite{bert18}. At the level 2.5 the joint distribution of all empirical densities 
defined in Eq. \eqref{defden} and all empirical currents defined in Eq. \eqref{defcur} is considered. In our notation $\rho(t)$ represents a vector with the empirical densities that has 
dimension $\Omega$ and $J(t)$ is a vector with the empirical currents  that has dimension $M$, where $M$ is the number of unordered pairs of states with non-zero transition rates.
The advantage of considering this level of large deviations is 
that  the rate function can be calculated exactly as  
\begin{equation}
I^{\textrm{cur}}_{2.5}\bigl[ (J(t))_{t\in [0,\tau]} ,(\rho(t))_{t\in [0,\tau] }\bigr]= \frac{1}{\tau}\int_0^{\tau}dt\sum_{i<j}\psi\left(J_{ij}(t),G_{ij}(t),a_{ij}(t)\right), 
\label{rate2.5b}
\end{equation}
where 
\begin{equation}\label{sonno1}
G_{ij}(t)\equiv \rho_i(t)w_{ij}(t)-\rho_j(t)w_{ji}(t),
\end{equation}
\begin{equation}\label{sonno2}
a_{ij}(t)\equiv 2\sqrt{\rho_i(t)\rho_j(t)w_{ij}(t)w_{ji}(t)},
\end{equation}
and
\begin{equation}
\psi(J,G,a)= \sqrt{G^2+a^2}-\sqrt{J^2+a^2}+J[\sinh^{-1}(J/a)-\sinh^{-1}(G/a)].
\end{equation}
Note that the quantities $G$ and $a$ depend on the empirical density $\rho$. The empirical density and current in Eq. \eqref{rate2.5b} fulfill the constraint 
\begin{equation}
\frac{d}{dt}\rho_i(t)+\sum_{j\neq i}J_{ij}(t)=0,
\label{constraintJ}
\end{equation}
for all states $i$. To simplify the notation we write $I^{\textrm{cur}}_{2.5} [J(t) ,\rho(t) ] $ instead of the l.h.s. of Eq. \eqref{rate2.5b}.

The name level 2.5 large deviations can also refer to the rate function associated with the 
joint probability of the empirical density and the empirical flow defined in Eq. \eqref{defflow}. The rate function with the empirical current can be obtained  
from the rate function with the empirical flow  \cite{bert18}.

An important technique in large deviation theory is the so called contraction \cite{elli85,demb98,holl00,touc09}, for which the rate function associated with a coarse-graining 
of the number of variables can be obtained from the original rate function. Hence, the rate function for an arbitrary current $J_\alpha$ can be obtained 
from a contraction of $I_{2.5}^{\textrm{cur}}[J(t),\rho(t)]$, which leads to the expression 
\begin{equation}
I_\alpha(x)=\inf_{ J(t) , \rho(t)  }  I_{2.5}^{\textrm{cur}}[J(t) ,\rho(t)]\,,\label{contracur}
\end{equation}
where $J(t)$ and $\rho(t)$ are such that they fulfill Eq. \eqref{constraintJ} and the relation 
\begin{equation}
\frac{1}{\tau}\int_{0}^\tau dt\sum_{i<j}\alpha_{ij}(t)J_{ij}(t)=x. 
\end{equation}
In particular, this relation leads to the inequality  
\begin{equation}
I_\alpha(x)\le I_{2.5}^{\textrm{cur}}[\tilde{J}(t),\tilde{\rho}(t)]= \frac{1}{\tau}\int_0^{\tau}dt\sum_{i<j}\psi\left(\tilde{J}_{ij}(t),\tilde{G}_{ij}(t),\tilde{a}_{ij}(t)\right),
\label{ineInonopt}
\end{equation}
where $\tilde{G}$ and $\tilde{a}$ are functions of  $\tilde{\rho}$ as in \eqref{sonno1} and \eqref{sonno2}. This inequality is valid for any pair of vectors that fulfill the constraints 
\begin{equation}
\frac{d}{dt}\tilde{\rho}_i(t)+\sum_{j\neq i}\tilde{J}_{ij}(t)=0,
\label{eqconstraint1}
\end{equation}
for all states $i$, and
\begin{equation}
\frac{1}{\tau}\int_0^{\tau}dt\sum_{i<j}\tilde{J}_{ij}(t)\alpha_{ij}(t)=x.
\label{eqconstraint2}
\end{equation}
The inequality \cite{ging16}
\begin{equation}
\psi\left(J_{ij} ,G_{ij} ,a_{ij}\right)\le \frac{1}{4}\frac{[J_{ij} -G_{ij} ]^2}{G_{ij} }\ln\frac{\rho_i w_{ij} }{\rho_j w_{ji}}
\label{gingine}
\end{equation}
together with Eq. \eqref{ineInonopt}, leads to
\begin{equation}
I_\alpha(x)\le \frac{1}{\tau}\int_0^{\tau}dt\sum_{i<j}\frac{1}{4}\frac{[\tilde{J}_{ij}(t)-\tilde{G}_{ij}(t)]^2}{\tilde{G}_{ij}(t)}\ln\frac{\tilde{\rho}_i(t)w_{ij}(t)}{\tilde{\rho}_j(t)w_{ji}(t)}. 
\label{boundrf}
\end{equation}

We are now left with the problem of finding a judicious choice of $\left(\tilde{J}(t),\tilde{\rho}(t)\right)$ that fulfills the constraints in 
Eq. \eqref{eqconstraint1} and in Eq. \eqref{eqconstraint2}. One such choice is
\begin{align}
	& \tilde{\rho}_i(t)= \pi_i(t) \label{testa} \\
	& \tilde{J}_{ij}(t)= \mathcal{J}_{ij}(t)+\frac{(x- \mathcal{J}_\alpha)K_{ij}}{\sum_{i'<j' }  K_{i'j'}\bar{\alpha}_{i'j'}},
	\label{eqchoice}
\end{align}
where 
\begin{equation}
\bar{\alpha}_{ij}\equiv\frac{1}{\tau}\int_0^\tau \alpha_{ij}(t)dt.
\label{alphaavg}
\end{equation}
The time-independent parameters $K_{ij}$ are antisymmetric, i.e., $K_{ij}=-K_{ji}$, and satisfy
\begin{equation}
\sum_{j\neq i}K_{ij}=0,
\label{constraintK}
\end{equation}
for all states $i$. 
Using this choice in Eq. \eqref{boundrf}, we obtain 
\begin{equation}
I_\alpha(x)\le \frac{\sigma_K^*}{4\mathcal{J}_K^2}(x-\mathcal{J}_\alpha)^2,
\label{mainresult2}
\end{equation}
where 
\begin{equation}\label{angelo}
\mathcal{J}_K\equiv \sum_{i<j} K_{ij}\bar{\alpha}_{ij}, 
\end{equation}
and
\begin{equation}
\sigma^*_K\equiv \frac{1}{\tau}\int_0^{\tau}\sum_{i<j}\frac{(K_{ij})^2}{ \mathcal{J}_{ij}(t)}\ln\frac{\pi_i(t)w_{ij}(t)}{\pi_j(t)w_{ji}(t)}dt. 
\label{sigmastardef}
\end{equation}
The global bound in Eq. \eqref{mainresult2}, together with Eq. \eqref{DIsecond}, leads to
\begin{equation}
D_\alpha\ge \frac{{\mathcal{J}}_K^2}{\sigma_K^*}.
\label{BoundD2}
\end{equation}
	
\subsection{Role of the parameter $K$}	
	
\subsubsection{Generic choice for $K$}
\label{sec421}

Due to the constraint in Eq. \eqref{constraintK}, $K_{ij}$ can be seen as the current of some auxiliary Markov process with 
time-independent transition rates in the stationary state. 
A natural choice of $K_{ij}$ is to consider the time-integrated probability current, as defined in Eq. \eqref{defcurrtime}, i.e., 
\begin{equation}\label{universo}\
K_{ij} \equiv \bar{\mathcal{J}}_{ij}.
\end{equation}
For this choice
\begin{equation}
\label{volo}
\mathcal{J}_K =  \sum_{i<j}\bar{\mathcal{J}}_{ij}\bar{\alpha}_{ij},
\end{equation}
and $\sigma^*_K=\sigma^*$, where $\sigma^*$ is defined in Eq.  \eqref{defsigmastar}. For currents with
time-independent increments $\alpha_{ij}(t)=\alpha_{ij}$,  we obtain $\sum_{i<j}\bar{\mathcal{J}}_{ij}\bar{\alpha}_{ij}=\mathcal{J}_\alpha$, where $\mathcal{J}_\alpha$
is given by Eq. \eqref{pranzo},
and  the bound in Eq. \eqref{mainresult2} becomes the bound in Eq. \eqref{mainresultspe1}. For currents with time-dependent increments, which include the 
rate of extracted work and the rate of heat flow in a heat engine driven by periodic temperature variation, the rate $\mathcal{J}_K$
in Eq. \eqref{volo} is, in general, different from the average current $\mathcal{J}_\alpha$.

\subsubsection{Other possible choices for $K$}	

The freedom of choice for the parameter $K$ depends on the network of states of the Markov process, with Eq. \eqref{constraintK} limiting  the number of independent currents 
$K_{ij}$ \cite{schn76}. For instance, for the unicyclic model in Fig. \ref{fig1a}, there is just one independent current  and $K_{ij}$ is the same 
for all pairs of states. In this case, the ratio $\sigma_K^*/\mathcal{J}_K^2$ becomes independent of $K$ and, therefore, there is only one bound in Eq. \eqref{mainresult2} 
regardless of the  value of $K_{ij}$. We note that the same argument about the freedom of choice for the parameter $K$ applies to stochastic protocols, as is the case 
of the model in Fig.\ref{fig1b} 

If we consider a model with the network of states shown in Fig. \ref{fig1c}, then there are two independent  $K_{ij}$ and different choices for these parameters can lead 
to different bounds  in Eq. \eqref{mainresult2}. Two particularly appealing choices for the parameter $K$ are the choices that conserve the rate of entropy production or 
the average current in Eq. \eqref{mainresult2}. The first choice corresponds to  a $K$  that fulfills the relation $\sigma_K^*=\sigma$ and the second choice corresponds 
to a $K$ that fulfills the relation  $\mathcal{J}_K=\mathcal{J}_\alpha$. Whether it is possible to set $K$ in such a way that one of these relations is fulfilled is a question that 
depends on the model (or class of models) at hand.

\subsection{Local bound }
\label{sec43}

We now derive a local quadratic bound on $I_\alpha(x)$ that leads to the first inequality in Eq. \eqref{boundD}.
For  $a$ and $G$ fixed, a  Taylor expansion of the function $\psi(J,G,a)$ for $J$ around the value $G$, leads to 
\begin{equation}
\psi(J,G,a)= \frac{(J-G)^2}{2\sqrt{G^2+a^2}}+\textrm{o}(|J-G|^2).
\label{taylorexp}
\end{equation}
Applying this Taylor expansion to Eq. \eqref{ineInonopt} with $\tilde{\rho}$ and $\tilde{J}$ given by \eqref{testa} and \eqref{eqchoice}, respectively, we obtain the local bound
\begin{equation}
\label{anguria}
I_\alpha(x)\le \frac{\tilde{\sigma}_K}{4\mathcal{J}_K^2}(x-\mathcal{J}_\alpha)^2+\textrm{o}(|x-\mathcal{J}_\alpha|^2),
\end{equation}
where $\mathcal{J}_K$ is   defined in Eq. \eqref{angelo} and 	
\begin{equation}
\tilde{\sigma}_K\equiv
\frac{1}{\tau}\int_{0}^{\tau}dt\sum_{i<j}\frac{2
(K_{ij})^2}{\pi_i(t) w_{ij}(t)+\pi_j(t) w_{ji}(t)}.
\label{sigmatildeK}
\end{equation}
The local bound in Eq. \eqref{anguria} together with Eq. \eqref{DIsecond} leads to
\begin{equation}\label{boundD2}
D_\alpha\ge \frac{ \mathcal{J}_K^2}{\tilde{\sigma}_K}  
\end{equation}
A generic model-independent choice for $K$ is the one given in Eq.  \eqref{universo}, i.e., $K_{ij}=\bar{\mathcal{J}}_{ij}$. 
If, in addition, the increments are time-independent, the bound in Eq. \eqref{boundD2} 
becomes the trade-off relation between speed and precision in Eq. \eqref{Fanotrade}. We recall that  from Eq. \eqref{basicine}, $\sigma^*_K\geq \tilde{\sigma}_K$, thus, the bound in Eq. \eqref{boundD2} is 
stronger than the bound in Eq. \eqref{BoundD2}.

\subsection{Bounds for time-independent transition rates}

Here, we stress that the bounds for time-periodic transition rates derived above imply new bounds for the case of time-independent transition
rates that lead to a non-equilibrium stationary state.  For time-independent transition rates, and for currents with time-independent increments, the terms in Eq. \eqref{mainresult2} become 
\begin{equation}
\mathcal{J}_K\equiv \sum_{i<j} K_{ij}\alpha_{ij}, 
\end{equation}
and
\begin{equation}
\sigma^*_K\equiv \sum_{i<j}\frac{(K_{ij})^2}{\mathcal{J}_{ij}}\ln\frac{\pi_i w_{ij}}{\pi_j w_{ji}}. 
\end{equation}
Hence, from Eq. \eqref{mainresult2} we have the bound 
\begin{equation}
I_\alpha(x)\le \frac{\sigma^{*}_K}{4\mathcal{J}_K^2}(x-\mathcal{J}_\alpha)^2.
\label{mainresult3}
\end{equation}
For $K_{ij}=\mathcal{J}_{ij}$, Eq. \eqref{mainresult3} becomes the known parabolic bound for stationary states from \cite{piet16,ging16}. 
Furthermore, for time-independent transition rates Eq. \eqref{boundD2} becomes
\begin{equation}
D_\alpha\ge \frac{\mathcal{J}_K^2}{\tilde{\sigma}_K},
\label{BoundD4}
\end{equation}
where 
\begin{equation}
\tilde{\sigma}_K\equiv
\sum_{i<j}\frac{2
(K_{ij})^2}{\pi_i w_{ij}+\pi_j w_{ji}}.
\end{equation}
This bound is tighter then the bound on the diffusion coefficient that follows from Eq. \eqref{mainresult3}. For the case $K_{ij}=\mathcal{J}_{ij}$,
Eq. \eqref{BoundD4} becomes an even stronger bound than the thermodynamic uncertainty relation, as discussed in Sec. \ref{sec3}.

\subsection{Second global bound}
\label{sec45}

We can obtain a bound different from the global bound in Eq. \eqref{mainresult2} by considering a choice for $\tilde{J}_{ij}(t)$ that is different from
the one in Eq. \eqref{eqchoice}. We write  the stationary distribution of a master equation with frozen transition rates  $w_{ij}(t)$ as $\mu_i(t)$.
This quantity is known as accompanying density \cite{hang82}. 
Due to the periodicity of $w_{ij}(t)$ we have $\mu_i(t)=\mu_i(t+\tau)$.
We consider the bound in Eq. \eqref{boundrf} with  $\tilde \rho_{i}(t)=\pi_i(t)$ and
\begin{equation}
\tilde J_{ij}\equiv\mathcal J_{ij}(t)+c_1(t)M_{ij}(t)+c_2(t) K_{ij}\,,
\label{eqSecondClassChoice}
\end{equation}
where $c_1(t)$ and $c_2(t)$ are time-periodic functions, $K_{i,j}$ is antisymmetric and fulfill the  relation in Eq. \eqref{constraintK}, and  
\begin{equation}
M_{ij}(t)\equiv\mu_i(t)w_{ij}(t)-\mu_j(t)w_{ji}(t)\,.
\end{equation}
Since $\sum_{j\neq i}M_{ij}(t)=0$, which comes from the definition of the accompanying density $\mu_i(t)$, this choice fulfills the constraint in Eq. \eqref{eqconstraint1}. 
Setting $K_{ij}=\bar{ \mathcal J}_{ij}$,
$c_1(t)=c_1$, and $c_2(t)=c_2$, the constraint in Eq. \eqref{eqconstraint2} applied to the choice in Eq. \eqref{eqSecondClassChoice}, leads to 
\begin{align}
&c_1=\left(x-\mathcal J_\alpha\right)q\mathcal J_\mu^{-1}\\
&c_2=\left(x-\mathcal J_\alpha\right)(1-q)\left(\sum_{i<j}\bar{\mathcal J}_{ij}\bar{\alpha}_{ij}\right)^{-1}.
\end{align}
where $q$ is an arbitrary real number and 
\begin{equation}
\mathcal J_\mu\equiv\frac{1}{\tau}\int_0^\tau\sum_{i<j}\alpha_{ij}(t)M_{ij}(t)dt.
\end{equation}
The bound in Eq. \eqref{boundrf} then becomes 
\begin{align}\label{unbound}
& I_\alpha(x)\le \frac{(x-\mathcal J_\alpha)^2}{4\tau}\int_0^\tau
\sum_{i<j}\frac{\left(M_{ij}(t)q\mathcal J_\mu^{-1}+(1-q)\left(\sum_{i<j}\bar{\mathcal J}_{ij}\bar{\alpha}_{ij}\right)^{-1}\bar{\mathcal J}_{ij}\right)^2}{\mathcal J_{ij}(t)}\log\frac{\pi_i(t)w_{ij}(t)}{\pi_j(t)w_{ji}(t)}dt
\end{align}
Minimization over the single parameter $q$ gives the tightest bound on the large deviation function. 
For $q=0$ we obtain the  bound in Eq. \eqref{mainresult2} with $K_{ij}=\bar{\mathcal J}_{ij}$. However, 
for $q=1$ we obtain a bound that cannot be obtained from Eq. \eqref{mainresult2}, which reads
\begin{equation}
I_\alpha(x)\leq \frac{\sigma^*_{\mu}}{4\mathcal J_\mu^2}(x-\mathcal J_\alpha)^2
\end{equation} 
where
\begin{equation}
\sigma^*_{\mu}=\frac{1}{\tau}\int_0^\tau
\sum_{i<j}\frac{\left(M_{ij}(t)\right)^2}{\mathcal J_{ij}(t)}\log\frac{\pi_i(t)w_{ij}(t)}{\pi_j(t)w_{ji}(t)}dt.
\end{equation}


\section{Conclusion}
\label{sec5}

The thermodynamic uncertainty relation and the parabolic bound on current fluctuations that generalizes it, constitute major recent developments 
in stochastic thermodynamics that are valid for Markov processes with time-independent transition rates that reach a stationary state, 
which describes a system driven by fixed thermodynamic forces. We have generalized these bounds to periodically driven systems. Similar to the 
bound for stationary states, we obtained a bound that depends on the single average rate $\sigma^*$ and on the average current. However,
for periodically driven systems this average rate is, in general, different from the thermodynamic entropy production $\sigma$. 
These rates have two essential physical properties: while $\sigma$ quantifies the energetic cost of maintaining the system out of 
equilibrium,  $\sigma^*$ provides a generic limit to current fluctuations.

The quite high degree of universality  of our results are encouraging with respect to possible applications. For instance, we have found a 
trade-off relation between speed and precision in periodically driven systems for currents that have time-independent increments. Physically, such relation
tells us that if one wants to generate net motion in a artificial molecular pump driven by an external periodic protocol, there is a universal  
limit on how fast and precise this net motion can be. 

For the case of the thermodynamic uncertainty relation for stationary states, several applications have been proposed \cite{bara15,piet16b,piet18,nguy16}. 
Figuring out how to extend these applications to periodically driven systems is an interesting direction for future work. One particular 
instance would be to extend the universal relation between power, efficiency and fluctuations from \cite{piet18}
to periodically driven heat engines. The more general bounds derived in Sec. \ref{sec4} that apply to time-dependent increments, 
might be important for these applications. Finally, good candidates for an experimental observation of the bounds we have derived here are periodically driven colloidal particles
and artificial molecular pumps.

\appendix

\section{Stochastic Protocol}
\label{appa}

\subsection{Mathematical definitions}
The master equation for the model with a stochastic protocol reads 
\begin{equation}
\frac{d}{dt} P_i^n=\sum_{j\neq i}\left(P_j^nw_{ji}^n-P_i^nw_{ij}^n\right)+\gamma(P_i^{n-1}-P_i^{n}), 
\label{mastereqsto}
\end{equation}
where $n-1=N-1$ for $n=0$ and $P_i^n$ is the time-dependent distribution. The stationary  distribution of state $(i,n)$ is denoted by $\pi_i^n$.
The stationary distribution of the state $n$ of the protocol is given by $\pi^n\equiv \sum_i\pi_i^n=1/N$, which comes from the solution of the 
master equation \eqref{mastereqsto} for the stationary distribution. The conditional probability for the system to be in state $i$ given that the protocol is in state 
$n$ is written as $\pi(i|n)= \pi_i^n/\pi^n=N\pi_i^n$. Consider a time-periodic Markov process with rates $w_{ij}(t)$ and period $\tau$. If the transition 
rates fulfill the relation $w_{ij}^n=w_{ij}(t=n\tau/N)$ and $\gamma=N/\tau$, then, in the limit $N\to \infty$, $\pi(i|n)\to \pi_i(t)$ \cite{bara16}, 
where $n= \left[tN/\tau\right]$ and $\left[\cdot\right]$ denotes the integer part. 
Therefore, if we consider the average elementary current $\mathcal{J}_{ij}^n\equiv\pi_i^nw_{ij}^n-\pi_j^nw_{ji}^n$ in the limit of $N\to \infty$,
we obtain
\begin{equation}
N\mathcal{J}_{ij}^n\to \mathcal{J}_{ij}(t),
\label{conectform}
\end{equation}
where $n= \left[tN/\tau\right]$. This relation is important for the connection between the cases of a deterministic and stochastic protocols.

A stochastic trajectory is denoted by $(b_t)_{0\le t\le t_f}$, where $t_f$ is the final time. Note that a state of the Markov process here is specified  
by the variable that determines the state of the system $i$ and the variable that determines the state of the protocol $n$. The stochastic 
trajectory has a fluctuating number of jumps $N_f$, the time interval between two jumps is denoted $\Delta t_k$, with $k=0,1,\ldots,N_f$, and 
the state of the Markov process during the time-interval $\Delta t_k$ is denoted $b_k$. 

The empirical density of state $(i,n)$, which is the fraction of time spent in this state, is defined as   
\begin{equation}
\rho_i^n= \frac{1}{t_f}\sum_{k=0}^{N_f} \Delta t_k \delta_{b_k,(i,n)}, 
\label{conectforma}
\end{equation}
$\delta_{b_k,(i,n)}$ is the Kronecker delta between the state of the trajectory $b_k$ and the state $(i,n)$. The notation here in the appendix is different 
from the notation in the main text for the case of a deterministic protocol. If we compare Eq. \eqref{conectforma} with Eq. \eqref{defden}, we see that here 
the upper index in $\rho_i^n$ refers to the state of the stochastic protocol and is equivalent to $t$ in $\rho_i^{(m)}(t)$, for which the upper index $m$ 
refers to the time interval of the stochastic trajectory. For a more compact notation we do not keep the dependence of the fluctuating quantities on the 
time interval $t_f$.
 
The empirical current from state $(i,n)$ to state $(j,n)$ reads 
\begin{equation}
J_{ij}^n= \frac{1}{t_f}\sum_{k=1}^{N_f}\left( \delta_{b_{k-1},(i,n)}\delta_{b_k,(j,n)}-\delta_{b_{k-1},(j,n)}\delta_{b_k,(i,n)}\right).
\label{conectform2}
\end{equation}
For the case of a stochastic protocol, we also consider the empirical flow (or unidirectional current) from state $(i,n)$ to state $(i,n+1)$, where $n+1=0$ for $n=N-1$, which is 
defined as
\begin{equation}
C_{i}^n= \frac{1}{t_f}\sum_{k=1}^{N_f} \delta_{b_{k-1},(i,n)}\delta_{b_k,(i,n+1)}.
\label{conectform3}
\end{equation}
The average of this empirical flow in the stationary state is $\mathcal{C}_i^n\equiv \langle C_i^n\rangle=\gamma \pi_i^n$.

A generic fluctuating current is written as  
\begin{equation}
J_\alpha\equiv \sum_{n=0}^{N-1}\sum_{i<j}\alpha_{ij}^nJ_{ij}^n,
\label{jalphasto}
\end{equation} 
where $\alpha_{ij}^n=-\alpha_{ji}^n$ are the increments. If we compare this expression with Eq. \eqref{defjalpha}, which 
is the expression for a deterministic protocol, we see 
that an integral over a period divided by the period $\tau$ for a deterministic protocol becomes a sum over $n$ divided 
by the total number of states of the protocol $N$ for a stochastic protocol. Note that the factor $1/N$ does not appear 
in front of the sum in the r.h.s of Eq. \eqref{jalphasto} due to Eq. \eqref{conectform}. The average current in the stationary state reads   
\begin{equation}
\mathcal{J}_\alpha\equiv\langle J_\alpha\rangle= \sum_{n=0}^{N-1}\sum_{i<j}\alpha_{ij}^n\mathcal{J}_{ij}^n. 
\end{equation} 

The rate function associated with $J_\alpha$ is defined as  
\begin{equation}
\textrm{Prob}(J_\alpha\approx x)\sim \exp[-t_f I_\alpha(x)],
\end{equation}
where $\sim$ means asymptotic equality in the limit $t_f\to\infty$. The scaled cumulant generating function for a 
stochastic protocol is defined as 
\begin{equation}
\lambda_\alpha(z)\equiv \lim_{t_f\to \infty}\frac{1}{t_f}\ln\langle\exp(t_f J_\alpha z)\rangle.
\label{defscaled2}
\end{equation}
These two quantities are related by a Legendre-Fenchel transform, as in Eq. \eqref{letransf}.

Similar to Eq. \eqref{defcurrtime} and Eq. \eqref{alphaavg} for a deterministic protocol, we define
\begin{equation}
\bar{\mathcal{J}}_{ij}\equiv \sum_{n=0}^{N-1}\mathcal{J}^n_{ij} 
\end{equation}
and
\begin{equation}
\bar{\alpha}_{ij}\equiv \frac{1}{N}\sum_{n=0}^{N-1}\alpha_{ij}^n, 
\end{equation}
respectively. Furthermore, we define 
\begin{equation}
\mathcal{J}_K\equiv \sum_{i<j} K_{ij}\bar{\alpha}_{ij}, 
\end{equation}
which is equivalent to \eqref{angelo},
\begin{equation}
\sigma^*_K\equiv \frac{1}{N^2}\sum_{n=0}^{N-1}\sum_{i<j}\frac{(K_{ij})^2}{\mathcal{J}_{ij}^n}\ln\frac{\pi_i^nw_{ij}^n}{\pi_j^{n}w_{ji}^n}, 
\end{equation}
which is equivalent to Eq. \eqref{sigmastardef}, and
\begin{equation}
\tilde{\sigma}_K\equiv \frac{1}{N^2}\sum_{n=0}^{N-1}\sum_{i<j}\frac{2(K_{ij})^2}{\pi_i^nw_{ij}^n+\pi_j^{n}w_{ji}^n},
\end{equation}
which is equivalent to Eq. \eqref{sigmatildeK}. The parameter $K_{ij}$ in these equations is anti-symmetric, i.e., $K_{ij}=-K_{ji}$, and thus 
fulfill $\sum_{j\neq i}K_{ij}=0$ for all $i$. 

\subsection{Proofs of the bounds}

We now consider the joint distribution of the vector of empirical densities $\rho$, the vector of
 empirical currents $J$, and the vector of the empirical flow $C$. The level 2.5 rate function \cite{bert15bis} 
for this Markov process reads 
\begin{equation}
I_{2.5}[J,C,\rho]= \sum_{n=0}^{N-1}\sum_{i<j}\psi\left(J_{ij}^n,G^{n}_{ij},a_{ij}^n\right)+\sum_{n=0}^{N-1}\sum_{i}\left(C_i^n\ln\frac{C_i^n}{\rho_i^n\gamma}+\gamma\rho_i^n-C_i^n\right), 
\end{equation}
where 
\begin{equation}
G^{n}_{ij}\equiv \rho_i^nw_{ij}^n-\rho_j^nw_{ji}^n,
\end{equation}
and
\begin{equation}
a_{ij}^n\equiv 2\sqrt{\rho_i^n\rho_j^nw_{ij}^nw_{ji}^n}.
\end{equation}
The quantities in this rate function fulfill the constraint
\begin{equation}
(C_i^{n}-C_i^{n-1})+\sum_{j\neq i}J_{ij}^n=0,
\label{eqconstraint3}
\end{equation}
for all $i$ and $n$.

Applying a contraction to obtain $I_\alpha(x)$ from $I_{2.5}[J,C,\rho]$, as in Eq. \eqref{contracur} for a deterministic protocol, and 
setting $\rho_i^n= \pi_i^n$ and $C_i^n= \gamma \pi_i^n$, we obtain 
\begin{equation}
I_\alpha(x)\le \sum_{n=0}^{N-1}\sum_{i<j}\psi\left(\tilde{J}_{ij}^n,\mathcal{J}^{n}_{ij},2\sqrt{\pi_i^n\pi_j^nw_{ij}^nw_{ji}^n}\right),
\label{boundrfsto}
\end{equation}
where $\tilde{J}_{ij}^n$ fulfill the constraints 
\begin{equation}
\sum_{n=0}^{N-1}\sum_{i<j}\tilde{J}_{ij}^n\alpha_{ij}^n=x
\label{eqconstraint2sto}
\end{equation}
and 
\begin{equation}
\gamma(\pi_i^{n}-\pi_i^{n-1})+\sum_{j\neq i}\tilde{J}_{ij}^n=0,
\label{eqconstraint4}
\end{equation}
for all $i$ and $n$.

The global bound on large deviations is obtained by setting 
\begin{equation}
\tilde{J}_{ij}^n= \mathcal{J}_{ij}^{n}+\frac{(x-\mathcal{J}_\alpha)K_{ij}}{\sum_{i<j}K_{ij}\bar{\alpha}_{ij}}.
\label{eqchoicesto}
\end{equation}
and by using the inequality in Eq. \eqref{gingine}. With these operations, Eq. \eqref{boundrfsto} becomes 
\begin{equation}
I_\alpha(x)\le \frac{\sigma^{*}_K}{4(\bar{\mathcal{J}}_K)^2}(x-\mathcal{J}_\alpha)^2,
\label{mainresultsto}
\end{equation}
which is the global bound for a stochastic protocol. 

The choice in Eq. \eqref{eqchoicesto} and the Taylor expansion in Eq. \eqref{taylorexp},
together with Eq. \eqref{boundrfsto} lead to the local bound
\begin{equation}
\label{anguria2}
I_\alpha(x)\le \frac{\tilde{\sigma}_K}{4\mathcal{J}_K^2}(x-\mathcal{J}_\alpha)^2+\textrm{o}(|x-\mathcal{J}_\alpha|^2).
\end{equation}
Using the relation \eqref{DIsecond} for the diffusion coefficient we obtain the bound
\begin{equation}
D_\alpha\ge \frac{\mathcal{J}_K^2}{\tilde{\sigma}_K}.
\end{equation}
The choice $K_{ij}=\bar{\mathcal{J}}_{ij}$ for a stochastic protocol leads to bounds similar to the bounds
discussed in Sec. \ref{sec421} for a deterministic protocol.

A bound similar to the bound in Eq. \eqref{unbound} for a stochastic protocol can be obtained by setting 
$\tilde \rho_{i}^n=\pi_i^n$ and
\begin{equation}
\tilde J_{ij}^n\equiv\mathcal J_{ij}^n+c_1M_{ij}^n+c_2\bar{ \mathcal J}_{ij}\,,
\label{eqSecondClassChoicesto}
\end{equation}
where
\begin{equation}
M_{ij}^n\equiv\mu_i^n w_{ij}^n-\mu_j^n w_{ji}^n\,,
\end{equation}
and $\mu_i^n$ is the solution of the stationary master equation  $\sum_{j\neq i}\left(\mu_i^n w_{ij}^n-\mu_j^n w_{ji}^n\right)=0$.
Defining 
\begin{equation}
\mathcal J_\mu\equiv\frac{1}{N}\sum_n\sum_{i<j}\alpha_{ij}^nM_{ij}^n,
\end{equation}
and setting 
\begin{align}
&c_1=\left(x-\mathcal J_\alpha\right)q\mathcal J_\mu^{-1}\\
&c_2=\left(x-\mathcal J_\alpha\right)(1-q)\left(\sum_{i<j}\bar{\mathcal J}_{ij}\bar{\alpha}_{ij}\right)^{-1},
\end{align}
leads to the fulfillment of the constraint in Eq. \eqref{eqconstraint2sto}. 
With this choice for $\tilde \rho_{i}^n$ and $\tilde J_{ij}^n$, the bound in Eq. \eqref{boundrfsto} becomes 
\begin{align}\label{unboundsto}
& I_\alpha(x)\le \frac{(x-\mathcal J_\alpha)^2}{4}\frac{1}{N^2}\sum_n
\sum_{i<j}\frac{\left(M_{ij}^nq\mathcal J_\mu^{-1}+(1-q)\left(\sum_{i<j}\bar{\mathcal J}_{ij}\bar{\alpha}_{ij}\right)^{-1}\bar{\mathcal J}_{ij}\right)^2}{\mathcal J_{ij}^n}\log\frac{\pi_i^nw_{ij}^n}{\pi_j^nw_{ji}^n}.
\end{align}
In particular, for $q=1$ we obtain 
\begin{equation}
I_\alpha(x)\leq \frac{(x-\mathcal J_\alpha)^2\sigma^*_{\mu}}{4\mathcal J_\mu^2}
\end{equation} 
where
\begin{equation}
\sigma^*_{\mu}=\frac{1}{N^2}\sum_n \sum_{i<j}\frac{\left(M_{ij}^n\right)^2}{\mathcal J_{ij}^n}\log\frac{\pi_i^nw_{ij}^n}{\pi_j^nw_{ji}^n}.
\end{equation}

\section*{References}

\end{document}